# THE CONTINUED IMPORTANCE OF HABITABILITY STUDIES

*A white paper submitted in response to the NAS call on Exoplanet Science Strategy*

**Lead Author:**

Ramses M. Ramirez

Earth-Life Science Institute, Tokyo Institute of Technology, Tokyo 152-8550

email: rramirez@elsi.jp

phone: 03-5734-2183

**Co-authors:**

Dorian S. Abbot (University of Chicago)

Vladimir Airapetian (NASA Goddard Space Flight Center and American University)

Yuka Fujii (Earth-Life Science Institute and Tokyo Institute of Technology)

Keiko Hamano (Earth-Life Science Institute and Tokyo Institute of Technology)

Amit Levi (Harvard University)

Tyler D. Robinson (Northern Arizona University)

Laura Schaefer (Arizona State University)

Eric T. Wolf (University of Colorado at Boulder)

Robin D. Wordsworth (Harvard University)

## INTRODUCTION

The habitable zone (HZ) is the hypothesized circumstellar region where standing bodies of liquid water could be stable on a planetary surface. The *classical* HZ [1] suggests that $CO_2$ and $H_2O$ are the key greenhouse gases for habitable exoplanets as they are on Earth. This implies that the climates of potentially habitable exoplanets are regulated via an Earth-like carbonate-silicate cycle (or equivalent) that maintains planetary habitability over Gyr timescales [1]. Such assumptions do not suggest that lifeforms elsewhere *must* be Earth-like or that they could even evolve under such conditions, however. For example, can life evolve on a planet with a 10-bar $CO_2$ atmosphere that is located near its host star's outer edge? Likewise, could the different radiation environment on a M-star planet produce life? Would any such life be "Earth-like"? Is the classical $CO_2$-$H_2O$ HZ really the best and only targeting tool that should be used to find potentially habitable planets?

Theoretical habitability studies will remain important for addressing such questions. Moreover, a proper theoretical understanding of habitability is necessary to guide and interpret future observations. In Section 1, we summarize recent habitability research, which includes updates to our understanding of habitable zones, ocean worlds, and magma oceans. Some potential research directions are given. We then argue that improving theoretical habitability studies will require 1-D and 3-D climate modelers to discuss and attempt to resolve recent differences in model results (Section 2). Technological improvements in imaging are *also* needed to further better such studies. Nevertheless, some key observations can be made now and in the near-future (Section 3). We conclude by summarizing the aforementioned points.

## SECTION 1: HABITABILITY

### 1a) Habitable zones: extensions in space

The outer edge of the HZ is extraordinarily sensitive to the allowable greenhouse gas combinations. For instance, the solar system's outer edge extends from ~1.7 to 10 AU if young planets can accrete hundreds to thousands of bars of hydrogen, $H_2$, from the protoplanetary disk[2]. Habitable conditions (e.g. surface conditions warm enough to support liquid water) for such planets located within the classical HZ may last a few to tens of millions of years [3], possibly longer at farther distances [2] or if biological feedbacks can regulate $H_2$ [4].

Moreover, $H_2$ can be volcanically-outgassed within a background $CO_2$ atmosphere, increasing the classical HZ width by ~ 50% or more [5]. If mantle conditions can remain reducing enough and/or $H_2$ escape rates to space are sufficiently slowed, habitable conditions within the classical HZ may last on Gyr timescales [5][6].

In the latter scenario, such secondary greenhouse gases can confound carbonate-silicate cycle predictions suggesting that $CO_2$ pressures for habitable planets should increase towards the outer edge [7]. This is because additional absorbers (like $H_2$) lower the $CO_2$ amounts needed to support habitable surface conditions[5].

Related concerns:

a. Would it be possible for life to subsist on a $H_2$-based form of photosynthesis [8]?

b. The effects of secondary greenhouse gases on HZ boundaries are just beginning to be explored. Other greenhouse gas combinations should also be assessed.

**1b) Habitable zones: changes with time**

Although most HZ studies have focused on the main-sequence phase of stellar evolution, the pre-main-sequence is crucial in discussions of planetary habitability. The habitability of M-star planets should be questioned for at least two reasons:

a. Such HZ planets (e.g. TRAPPIST-1) should have lost much - if not all - of their surface water inventory following a runaway greenhouse episode during the superluminous pre-main-sequence phase [9]–[11]unless they could have accreted a few tens % of their mass in water [12], [13].
b. Large fluxes of stellar radiation (X-ray and UV) and stellar winds could significantly erode exoplanetary atmospheres over geological timescales [14].

Thus, M-star planets located in the HZ today may face extra challenges to their habitability. The entire spectrum of A – M stars should be observed with these caveats in mind.

Related concerns:

a. Can desiccated M-dwarf HZ planets be partially replenished by later meteoritic impacts or mantle degassing and become potentially habitable [9]?

**1c) Magma oceans as keys to understanding atmospheric evolution and habitability**

In addition, the pre-main-sequence phase is crucial for understanding how worlds retain and lose volatiles during this stage of early accretion [15].

Moreover, high levels of abiotically-produced oxygen, $O_2$, incorrectly linked to life, may accumulate in magma ocean atmospheres [10]. However, the resulting magma ocean and impact-induced soil oxidation processes may be important $O_2$ sinks [16][17]. Thus, the magma ocean stage may explain differences in evolutionary histories between Earth and Venus and could elucidate the conditions necessary for the emergence of life on exoplanets.

Related concerns:

a. Most magma ocean studies to date have considered mostly $H_2O$- or $CO_2$-dominated atmospheres [15][16][18][19], but many atmospheric compositions are possible depending on impactor chemistry [20]. These should all be explored.
b. This early stage of planetary evolution could be the key to providing a more definitive HZ inner edge, improving our ability to target worlds that are likely to have sizeable water inventories [15].

**1d) The possible habitability of ocean worlds**

It has been argued that the carbonate-silicate cycle requires some land to operate, which suggests that ocean worlds (which have no land and possess total water inventories that are many times that of Earth's) may not be habitable [21]. However, sea ice enriched in clathrates should readily form within the resulting high $CO_2$ atmospheric environments predicted for such planets, providing conditions favorable for the initiation of life [22]. Plus, the carbonate-silicate cycle may be replaced by one in which net outgassing of $CO_2$ is balanced by a net ocean influx [13]. High $CO_2$ levels may also cool the stratosphere sufficiently to prevent a moist greenhouse [22].

## SECTION 2: CLIMATE COMMUNITY COLLABORATIONS

### 2a) 1-D and 3-D climate models as planetary habitability assessment tools

Until recently, virtually all exoplanet habitability studies had been performed using 1-D climate models. However, the types of climate models used to perform habitability assessments have significantly diversified over the past ~5 years. Exoplanet habitability studies are now routinely performed using a wide array of 1-D (including energy balance models) (e.g. [5][9][15][16][18][19][22]–[24]) and 3-D (e.g. [25][26][27][28][29][30][31][32][33]) climate models, with some studies employing both [34][35] or even hybrid approaches [36].

This trend has occurred because both 1-D and 3-D climate models have complementary advantages and disadvantages. 1-D models are computationally cheap, allowing quick exploration of parameter space, more complex radiative transfer and atmospheric chemistry, and are more easily coupled to other (e.g. interior, escape, stellar, photochemical) models. For instance, atmospheric and magma ocean processes can be readily coupled in 1-D models [16], allowing volatile retention and $O_2$ build-up to be quantified. Also, coupled climate-photochemical calculations are routinely performed [24].

In contrast, 3-D models calculate more atmospheric processes self-consistently than can be done with 1-D models, which have to make more simplifying assumptions regarding dynamics, relative humidity, and clouds. Thus, 3-D models are used to assess the subtleties of atmospheres in greater detail, which includes evaluating complex circulation patterns.

### 2b) The need for collaborations between 1-D and 3-D modelers

Unfortunately, there is currently a lack of information exchange between the 1-D and 3-D climate modeling communities. However, community interactions are necessary for addressing recent unresolved differences in model predictions such as:

a. Does relative humidity increase with surface temperature as 1-D models predict [37], [38]or can it decrease as suggested by some 3-D models [39]?
b. Is the moist greenhouse for M-star inner edge planets triggered at the low mean surface temperatures predicted in some 3-D models [28][31] or at the higher temperatures calculated in other 3-D and 1-D models [1][40]?
c. Do surface temperature inversions in very warm atmospheres truly occur as predicted in recent 3-D simulations [41]?
d. How valid is the 1-D moist adiabatic lapse rate assumption for planetary atmospheres [1][41]?

## SECTION 3: CURRENT LIMITATIONS AND SUGGESTED OBSERVATIONS

All models have to make simplifications, however. With 3-D models, this usually means tuning poorly-understood processes or parameterizing them based on how they operate on Earth. With 1-D models, this requires making assumptions about the relative humidity and lapse rate.

Unfortunately, such simplified approaches are almost certainly inadequate to study exoplanets. Plus, current observations are also woefully limited. For example, present observational capabilities are insufficient to characterize terrestrial HZ planets and their atmospheres in great detail, requiring modelers to "invent" atmospheric compositions and guess what the values for specific planetary parameters may be. Thus, major technological advances in observational techniques are necessary to produce data that is of sufficient quality to inform and substantially improve exoplanet climate models. In spite of these limitations, some examples of observations that can still be made today and in the near-future include:

a. Do M-star HZ planets have atmospheres? If they do, does $O_2$ build up? Does $O_2$ build up on HZ planets orbiting other star types? HabEx and LUVOIR could observe atmospheres for signs of $O_2$ buildup at VIS/NIR wavelengths. ELTs can also detect $O_2$ through multiple transits [42].
b. OST and JWST [43]could detect $O_3$ in $O_2$-rich atmospheres.
c. The ELTs may be able to detect thermal emission from magma oceans [44][45].
d. The ELTs may observe nearby M-stars for planets located in the pre-main-sequence HZ since they can work at small inner working angles [9].
e. High predicted $CO_2$ pressures for outer edge classical HZ planets lower scale heights and increase the difficulty to observe bioindicators in transmission spectroscopy. However, $H_2$ can increase atmospheric scale height in sufficiently high quantities, which improves bioindicator detection [5][46]. JWST could target the TRAPPIST-1 planets and LHS-1140b [47].
f. Ocean worlds with sufficient water may be deduced from other types of habitable planets. Lower computed densities for ocean worlds would distinguish them from drier ones, breaking the mass-radius degeneracy. Moreover, should such atmospheres be $CO_2$-rich [13], lower scale heights would distinguish them from worlds with H/He outer envelopes.

Once enough planetary atmospheres inside and outside the HZ are characterized, statistics will be available to test and improve model predictions [1], [7]. In turn, improved climate models would be used to provide better interpretations of observations.

## FINAL SUMMARY

As the direct links between theory (Section 1) and observations (Section 3) suggest, theoretical habitability studies will remain indispensable for observations. This is because such studies, which include understanding the planetary (e.g. atmospheric, geologic) and stellar processes[1] that make a planet more or less likely to support life, provide the roadmap for making proper observations. Continued work on habitability is essential even if life elsewhere happens to be exactly like it is on Earth- which would *still* have appeared very differently to an extraterrestrial observer examining our world at different points in geologic history [48].

---

[1] Although we have focused on climate modeling in this white paper, we also support more theoretical habitability studies in other areas (e.g. interiors, photochemistry, volatile delivery, stellar atmospheres, aeronomy, and atmospheric escape).

The likelihood of finding life can be increased by suspending some of our most cherished Earth-centric notions of habitability, whatever they may be (e.g. carbonate-silicate cycles, $CO_2$-$H_2O$ atmospheres, oxygen biochemistries), no matter how compelling they may seem, and considering assessing alternate scenarios as well. We do not know what extraterrestrial life may be like, so it is self-restricting to generalize life elsewhere solely based on a limited understanding about this planet. Healthy scientific speculation will be key to making progress given limited observations. Both 1-D and 3-D climate models will be essential for advancing theoretical habitability studies. Community partnerships aimed at resolving model differences will lead to improved exoplanet climate models, maximizing the utility of such theoretical habitability studies. Such efforts would ultimately lead to improved observational interpretations as well.

However, improvements to theoretical modeling efforts are currently stymied by the lack of detailed exoplanetary observations. Although drawing upon knowledge from various solar system bodies to infer exoplanetary processes has been a useful tactic [1][5][6][15][49] having access to better observations would lead to even bigger advances in our understanding. This situation would vastly improve with technological (e.g. engineering) advancements in observational techniques, including direct imaging. Even with current limitations, we have suggested some observations that can still be made with upcoming and next generation missions.

## REFERENCES

[1] J.F. Kasting, D. Whitmire, and R. Raynolds, “Habitable Zones Around Main Sequence Stars,” *Icarus*, vol. 101, pp. 108–128, 1993.

[2] R. Pierrehumbert and E. Gaidos, “Hydrogen Greenhouse Planets Beyond the Habitable Zone,” *Astrophys. J. Lett.*, vol. 734, no. L13, Apr. 2011.

[3] R.D. Wordsworth, “Transient conditions for biogenesis on low-mass exoplanets with escaping hydrogen atmospheres,” *Icarus*, vol. 219, no. 1, pp. 267–273, 2012.

[4] D. S. Abbot, “A Proposal for Climate Stability on H2-greenhouse Planets,” *Astrophys. J. Lett.*, vol. 815, no. 1, 2015.

[5] R. M. Ramirez and L. Kaltenegger, “A Volcanic Hydrogen Habitable Zone,” *Astrophys. J. Lett.*, vol. 837, no. 1, 2017.

[6] R. M. Ramirez, R. Kopparapu, M. E. Zugger, T. D. Robinson, R. Freedman, and J. F. Kasting, “Warming early Mars with CO2 and H2,” *Nat. Geosci.*, vol. 7, no. 1, pp. 59–63, Nov. 2014.

[7] E. M.-R. Bean, Jacob L.; Abbot, Dorian S.; Kempton, “A Statistical Comparative Planetology Approach to the Hunt for Habitable Exoplanets and Life Beyond the Solar System,” *Astrophys. J. Lett.*, vol. 841, no. 2, 2017.

[8] W. Bains, S. Seager, and A. Zsom, “Photosynthesis in Hydrogen-Dominated Atmospheres,” pp. 716–744, 2014.

[9] R. M. Ramirez and L. Kaltenegger, “The habitable zones of pre-main-sequence stars,” *Astrophys. J. Lett.*, vol. 797, no. 2, 2014.

[10] R. Luger and R. Barnes, “Extreme water loss and abiotic O2 buildup on planets throughout the habitable zones of M dwarfs,” *Astrobiology*, vol. 15, no. 2, pp. 119–143, 2015.

[11] F. Tian and S. Ida, “Water contents of Earth-mass planets around M dwarfs,” *Nat.*

*Geosci.*, vol. 8, no. 3, 2015.

[12] Y. Alibert and W. Benz, "Formation and composition of planets around very low mass stars," *Astron. Astrophys.*, vol. 598, no. L5, 2017.

[13] M. Levi, A.; Sasselov, D.; Podolak, "The Abundance of Atmospheric CO2 in Ocean Exoplanets: a Novel CO2 Deposition Mechanism," *Astrophys. J.*, vol. 838, no. 1, 2017.

[14] M. W. Airapetian, Vladimir S.; Glocer, Alex; Khazanov, George V.; Loyd, R. O. P.; France, Kevin; Sojka, Jan; Danchi, William C.; Liemohn, Michael W.Airapetian, Vladimir S.; Glocer, Alex; Khazanov, George V.; Loyd, R. O. P.; France, Kevin; Sojka, Jan; Danchi, Wi, "How Hospitable Are Space Weather Affected Habitable Zones? The Role of Ion Escape," *Astrophys. J. Lett.*, vol. 836, no. 1, 2017.

[15] K. Hamano, Y. Abe, and H. Genda, "Emergence of two types of terrestrial planet on solidification of magma ocean," *Nature*, vol. 497, no. 7451, pp. 607–610, 2013.

[16] D. Schaefer, L., Wordsworth, R. D., Berta-Thompson, Z., Sasselov, "Predictions of the atmospheric composition of GJ 1132b," *Astrophys. J.*, vol. 2, no. 63, 2016.

[17] K. Kurosawa, "Impact-driven planetary desiccation: The origin of the dry Venus," *Earth Planet. Sci. Lett.*, vol. 429, no. 1, pp. 181–190, 2015.

[18] A. Marcq, E.; Salvador, A.; Massol, H.; Davaille, "Thermal radiation of magma ocean planets using a 1-D radiative-convective model of H2O-CO2 atmospheres," *J. Geophys. Res. Planets*, vol. 122, no. 7, pp. 1539–1553, 2017.

[19] E. Salvador, A.; Massol, H.; Davaille, A.; Marcq, E.; Sarda, P.; Chassefière, "The relative influence of H2O and CO2 on the primitive surface conditions and evolution of rocky planets," *J. Geophys. Res. Planets*, vol. 122, no. 7, 2017.

[20] L. Schaefer and B. Fegley Jr., "Chemistry of atmospheres formed during accretion of the Earth and other terrestrial planets," *Icarus*, vol. 208, no. 1, pp. 434–448, 2010.

[21] F. J. Abbot, Dorian S.; Cowan, Nicolas B.; Ciesla, "Indication of Insensitivity of Planetary Weathering Behavior and Habitable Zone to Surface Land Fraction," *Astrophys. J.*, vol. 756, no. 2, 2012.

[22] R. D. Wordsworth and R. T. Pierrehumbert, "Water Loss From Terrestrial Planets With Co 2 -Rich Atmospheres," *Astrophys. J.*, vol. 778, no. 2, p. 154, Dec. 2013.

[23] A. Vladilo, Giovanni; Silva, Laura; Murante, Giuseppe; Filippi, Luca; Provenzale, "Modeling the Surface Temperature of Earth-like Planets," *Astrophys. J.*, vol. 804, no. 1, 2015.

[24] K. Engin, J. L. Grenfell, M. Godolt, S. Barbara, and R. Heike, "The Effect of Varying Atmospheric Pressure upon Habitability and Biosignatures of Earth-like Planets," *Astrobiology*, vol. 18, no. 2, 2018.

[25] J. Yang, N. Cowan, and D. Abbot, "Stabilizing cloud feedback dramatically expands the habitable zone of tidally locked planets," *Astrophys. J. Lett.*, vol. 771, no. 2, 2013.

[26] A. P. Kaspi, Yohai; Showman, "Atmospheric Dynamics of Terrestrial Exoplanets over a Wide Range of Orbital and Atmospheric Parameters," *Astrophys. J.*, vol. 804, no. 1, 2015.

[27] S. Kopparapu, Ravi kumar; Wolf, Eric T.; Haqq-Misra, Jacob; Yang, Jun; Kasting, James F.; Meadows, Victoria; Terrien, Ryan; Mahadevan, "The Inner Edge of the Habitable Zone for Synchronously Rotating Planets around Low-mass Stars Using General Circulation Models," *Astrophys. J.*, vol. 819, no. 1, 2016.

[28] D. S. Fujii, Yuka; Del Genio, Anthony D.; Amundsen, "NIR-driven Moist Upper Atmospheres of Synchronously Rotating Temperate Terrestrial Exoplanets," *Astrophys. J.*, vol. 848, no. 2, 2017.

[29] G. Turbet, Martin; Forget, Francois; Leconte, Jeremy; Charnay, Benjamin; Tobie, “CO2 condensation is a serious limit to the deglaciation of Earth-like planets,” *Earth Planet. Sci. Lett.*, vol. 476, pp. 11–21, 2017.

[30] E. T. Wolf, “Assessing the Habitability of the TRAPPIST-1 System Using a 3D Climate Model,” *Astrophys. J. Lett.*, vol. 839, no. 1, 2017.

[31] R.K. Kopparapu; Wolf, Eric T.; Arney, Giada; Batalha, Natasha E.; Haqq-Misra, Jacob ; Grimm, Simon L.; Heng, “Habitable moist atmospheres on terrestrial planets near the inner edge of the habitable zone around M dwarfs,” *Astrophys. J.*, vol. 1, no. 5, 2017.

[32] J. D. Haqq-Misra, E. T. Wolf, M. Josh, X. Zhang, and R. K. Kopparapu, “Demarcating Circulation Regimes of Synchronously Rotating Terrestrial Planets within the Habitable Zone,” *Astrophys. J.*, vol. 852, no. 2, 2018.

[33] Y. Kodama, T.; Nitta, A.; Genda, H.; Takao, Y.; O’ishi, R.; Abe-Ouchi, A.; Abe, “Dependence of the Onset of the Runaway Greenhouse Effect on the Latitudinal Surface Water Distribution of Earth-Like Planets,” *J. Geophys. Res. Planets*, vol. 123, no. 2, pp. 559–574, 2018.

[34] D. S. Mills, Sean M.; Abbot, “Utility of the Weak Temperature Gradient Approximation for Earth-like Tidally Locked Exoplanets,” *Astrophys. J. Lett.*, vol. 774, no. 2, 2013.

[35] J. Checlair, K. Menou, and D. S. Abbot, “No snowball on habitable tidally locked planets,” *Astrophys. J.*, vol. 845, no. 2, 2017.

[36] L. Carone, R. Keppens, L. Decin, and T. Henning, “Stratosphere circulation on tidally locked ExoEarths,” *Mon. Not. R. Astron. Soc.*, vol. 473, no. 4, pp. 4672–4685, 2017.

[37] C. Goldblatt, T. D. Robinson, K. J. Zahnle, and D. Crisp, “Low simulated radiation limit for runaway greenhouse climates,” *Nat. Geosci.*, vol. 6, no. 8, pp. 661–667, Jul. 2013.

[38] R. M. Ramirez, R. K. Kopparapu, V. Lindner, and J. F. Kasting, “Can increased atmospheric CO2 levels trigger a runaway greenhouse?,” *Astrobiology*, vol. 14, no. 8, pp. 714–31, Aug. 2014.

[39] G. L. Russell, A. a. Lacis, D. H. Rind, C. Colose, and R. F. Opstbaum, “Fast atmosphere-ocean model runs with large changes in CO 2,” *Geophys. Res. Lett.*, vol. 40, no. 21, pp. 5787–5792, Nov. 2013.

[40] J. Leconte, F. Forget, B. Charnay, R. Wordsworth, and A. Pottier, “Increased insolation threshold for runaway greenhouse processes on Earth-like planets.,” *Nature*, vol. 504, no. 7479, pp. 268–71, Dec. 2013.

[41] E. T. Wolf and O. B. Toon, “Journal of Geophysical Research : Atmospheres the brightening Sun,” *J. Geophys. Res. Atmos.*, vol. 120, no. 12, pp. 5775–5794, 2015.

[42] F. Rodler and M. Lopez-Morales, “Feasibility studies for the detection of O2 in an Earth-like exoplanet,” *Astrophys. J.*, vol. 781, no. 1, 2014.

[43] J. K. Barstow and P. G. J. Irwin, “Habitable worlds with JWST: transit spectroscopy of the TRAPPIST-1 system?,” *Mon. Not. R. Astron. Soc.*, vol. 461, no. 1, pp. L92–L96, 2016.

[44] E. Miller-Ricci, M. R. Meyer, S. Seager, and L. Elkins-Tanton, “On the emergent spectra of hot protoplanet collision afterglows,” *Astrophys. J.*, vol. 704, no. 1, 2009.

[45] K. Hamano, H. Kawahara, Y. Abe, M. Onishi, and G. L. Hashimoto, “Lifetime and spectral evolution of a magma ocean with a steam atmosphere:its detectability by future direct imaging,” *Astrophys. J.*, vol. 806, no. 2, 2015.

[46] R. Seager, S.; Bains, W.; Hu, “Biosignature Gases in H2-dominated Atmospheres on Rocky Exoplanets,” *Astrophys. J.*, vol. 777, no. 2, 2013.

[47] C. V. Morley, L. Kreidberg, Z. Rustamkulov, T. D. Robinson, and J. J. Fortney,

"Observing the Atmospheres of Known Temperate Earth-sized Planets with JWST," *Astrophys. J.*, vol. 850, no. 2, 2017.

[48] L. Kaltenegger; Traub, Wesley A.; Jucks, "Spectral Evolution of an Earth-like Planet," *Astrophys. J.*, vol. 658, no. 1, pp. 598–616, 2007.

[49] T. D. Robinson and D. C. Catling, "Common 0.1 bar tropopause in thick atmospheres set by pressure-dependent infrared transparency," *Nat. Geosci.*, vol. 7, no. 1, pp. 12–15, Dec. 2013.